\documentstyle[aps,prl,multicol,psfig]{revtex} 
\newcommand{\comment}[1]{\vspace{-3 mm}\par 
\marginpar{\large\underline{}}\noindent
\framebox{\begin{minipage}[c]{0.465 \textwidth}
\rm #1 \end{minipage}}\vspace{1 mm}\par}
\def\mathbf#1{{\bf #1}}
\def\mathsf#1{{\sf #1}}
\def\mathbb#1{\mathbf{ #1}}
\def\mathfrak#1{\mathbf{ #1}}
\begin{document}

\title{
%
Variational Principles of
Lagrangian Averaged Fluid Dynamics
}
\author{
Darryl D. Holm
\\Theoretical Division and Center for Nonlinear Studies
\\Los Alamos National
Laboratory, MS B284
\\ Los Alamos, NM 87545
\\ email: dholm@lanl.gov
}

\maketitle

\begin{abstract}
The Lagrangian average (LA) of the ideal fluid equations preserves
their fundamental transport structure. This transport structure is responsible
for the Kelvin circulation theorem of the LA flow and, hence, for its
potential vorticity convection and its helicity conservation.

Lagrangian averaging also preserves the Euler-Poincar\'e (EP)
variational framework that implies the LA fluid equations. This is
expressed in the Lagrangian-averaged Euler-Poincar\'e (LAEP) theorem
proven here and illustrated for the Lagrangian average Euler (LAE)
equations. 
\\



\end{abstract} 

\vskip 4mm

\begin{multicols}{2}


\noindent
{\bf Introduction.}
In turbulence, in climate modeling and in other multiscale
fluids problems, a major challenge is ``scale-up.'' This is the
challenge of deriving models that correctly capture the mean, or
large scale flow -- including the influence on it of the rapid,
or small scale dynamics. 

Meteorology and oceanography must deal with averaging in either the
Eulerian, or the Lagrangian fluid specification. Determining the
relation between averaged quantities in these two specifications is one
of the classical problems in the physics of fluids.
 

The GLM (Generalized Lagrangian Mean) equations of Andrews {and}
McIntyre \cite{AM[1978]} systematize the approach to Lagrangian fluid
modeling by introducing a slow + fast decomposition of the Lagrangian
particle trajectory in general form. In these (exact, but not
closed) equations, the Lagrangian mean of a fluid quantity evaluated at
the mean particle position is related to its Eulerian mean,
evaluated at the {\it current} particle position. Thus, the GLM
equations are expressed directly in the Eulerian representation. The Lagrangian
mean has the advantage of preserving the fundamental {\bf transport
structure} of fluid dynamics. For example, the Lagrangian mean commutes
with the scalar advection operator and it preserves the Kelvin
circulation property of the fluid motion equation. However, the
Lagrangian mean also has two main {\it disadvantages}: it is history
dependent and it does not commute with the spatial gradient. 


In this paper, we place Lagrangian averaged (LA) fluid equations such as
the GLM equations into the Euler-Poincar\'e (EP) framework of
constrained variational principles. This demonstrates the
variational reduction property of the Lagrangian mean, encapsulated in
the LAEP Theorem proven here:
\comment{
{\bf Lagrangian Averaged Euler-Poincar\'e Theorem}
{\it Lagrangian averaging preserves the variational structure
of the Euler-Poincar\'e framework for fluids.}}

According to this theorem, the Lagrangian mean's preservation
of the fundamental transport structure of fluid dynamics also
extends to preserving its EP variational structure
\cite{HMR[1998a],HMR[1998b],MR[1999]}. This preservation of
variational structure is not possible with the Eulerian mean. The
Eulerian mean also does {\it not} preserve the transport structure of
fluid mechanics.

The LAEP Theorem puts the approach using LA Hamilton's
principles and LA equations onto equal footing. This is quite a bonus
for both approaches to modeling fluids. According to the LAEP Theorem,
the LA Hamilton's principle produces dynamics that is guaranteed
to be verified directly by averaging the original equations, and the
LA equations inherit the conservation laws that are available from the
symmetries of Hamilton's principle for fluids.

\noindent
{\bf Outline of the paper.} 
We begin by briefly reviewing the GLM theory of Andrews {and} McIntyre
\cite{AM[1978]}. We then state and prove the LAEP theorem, following the EP
framework of
\cite{HMR[1998a],HMR[1998b],MR[1999]}. We illustrate the LAEP theorem by
applying it to incompressible ideal fluids. Finally, we discuss recent progress
toward closure of these equations as models of fluid turbulence.

\noindent
{\bf GLM (Generalized Lagrangian Mean) theory.} The GLM theory
\cite{AM[1978]} begins by assuming that the Lagrange-to-Euler map {\it
factorizes} as a product of diffeomorphisms,
$
g(t)=\Xi\,(t)\cdot \tilde{g}(t)
\,.
$ 
Moreover, the first factor $\tilde{g}(t)$ arises from an averaging
process denoted as
$
\bar{g}(t)
=
\overline{\Xi\,(t)\cdot \tilde{g}(t)}
=
\tilde{g}(t)
\,.
$ 
The averaging process $\overline{(\,\cdot\,)}$ can be reasonably
arbitrary, except that it must satisfy the {\bf projection property},
so that 
$\bar{\bar{g}}(t)=\bar{\tilde{g}}(t)=\tilde{g}(t)$. 
Hence, a fluid parcel labeled by $\mathbf{x}_0$ has 
current position,
\begin{eqnarray*}
\mathbf{x}^\xi(\mathbf{x}_0,t)
\equiv
g(t)\!\cdot\!\mathbf{x}_0
=
\Xi\,(t)\!\cdot(\tilde{g}(t)\!\cdot\!\mathbf{x}_0)
=
\Xi\,(\mathbf{x}(\mathbf{x}_0,t),t)
\,,
\end{eqnarray*}
and it has mean position 
$ 
\mathbf{x}(\mathbf{x}_0,t)
=
\tilde{g}(t)\!\cdot\!\mathbf{x}_0
\,.
$ 
%

\noindent
{\bf GLM velocities and advective derivatives.} 
The composition of maps $g(t)=\Xi\,(t)\cdot \tilde{g}(t)$ yields via
the chain rule the following {\bf velocity relation},
\begin{equation}\label{vee-rel}
\dot{\mathbf{x}}^{\,\xi}(\mathbf{x}_0,t)
=
\dot{g}(t)\!\cdot\!\mathbf{x}_0
=
\dot{\Xi}\,(t)\!\cdot\!\mathbf{x}
+
T\Xi\cdot\!(\dot{\tilde{g}}(t)\!\cdot\!\mathbf{x}_0)
\,.
\end{equation}
By invertibility, 
$\mathbf{x}_0
=
g^{-1}(t)\!\cdot\!\mathbf{x}^\xi
=
\tilde{g}^{-1}(t)\!\cdot\!\mathbf{x}
\,.$ Hence, one may define the fluid parcel velocity at the current
position $\mathbf{u}(\mathbf{x}^\xi,t)$ in terms of a vector field
evaluated at the mean position $\mathbf{u}^\xi(\mathbf{x},t)$ as
\begin{equation}
\mathbf{u}(\mathbf{x}^\xi,t)
=
\dot{g}\cdot g^{-1}(t)\cdot\mathbf{x}^\xi
=
\dot{g}\cdot \tilde{g}^{-1}(t)\cdot\mathbf{x}
\equiv
\mathbf{u}^\xi(\mathbf{x},t)
\,.
\nonumber
\end{equation}
The velocity relation (\ref{vee-rel}) then implies, 
\begin{equation}\label{u-xi-def}
\mathbf{u}^\xi(\mathbf{x},t)
=
\frac{\partial \Xi}{\partial t} \,(\mathbf{x},t)
+
\frac{\partial \Xi}{\partial \mathbf{x}}
\cdot
\bar{\mathbf{u}}^L(\mathbf{x},t)
\,.
\end{equation}
This is a standard velocity relation from GLM theory, in which the {\bf
Lagrangian mean velocity} $\bar{\mathbf{u}}^L$ is {\it defined} as
\begin{equation}\label{LA-vee}
\bar{\mathbf{u}}^L(\mathbf{x},t)
\equiv
\overline{\mathbf{u}^\xi(\mathbf{x},t)}
=
\overline{\dot{g}\tilde{g}^{-1}(t) }\cdot \mathbf{x}
=
\dot{\tilde{g}}(t)\tilde{g}(t)^{-1}
\cdot
\mathbf{x}
\,.
\end{equation}
In the third equality one invokes the projection property of the
averaging process and finds
$\overline{\dot{g}}=\dot{\overline{g}}=\dot{\tilde{g}}$ from equation
(\ref{vee-rel}), so that
$ 
\bar{\mathbf{u}}^L(\mathbf{x},t)
=
\dot{\tilde{g}}(t)\tilde{g}(t)^{-1}
\cdot
\mathbf{x}
\equiv
\tilde{\mathbf{u}}(\mathbf{x},t)
\,.
$ 
Thus, the Lagrangian mean velocity $\bar{\mathbf{u}}^L$
coincides with $\tilde{\mathbf{u}}$, the tangent to the mean motion
associated with $\tilde{g}(t)$. Hence, one may write equation
(\ref{u-xi-def}) in terms of the  {\bf LA material time derivative}
$D^L/Dt$ as
\begin{equation}\label{u-Xi-rel}
\mathbf{u}^\xi(\mathbf{x},t)
=
\Big(\frac{\partial}{\partial t}
+
\bar{\mathbf{u}}^L\cdot\nabla\Big)\,\Xi(\mathbf{x},t)
\equiv
\frac{D^L}{Dt}\,\Xi\,(\mathbf{x},t)
\,.
\end{equation}

For any other fluid quantity $\chi$ one may similarly define
$\chi^{\,\xi}$ as the composition of functions
$ 
\chi^{\,\xi}(\mathbf{x},t)
=
\chi(\mathbf{x}^\xi,t)
=
\chi(\Xi\,(\mathbf{x},t),t)
.
$ 
Taking the LA material time derivative of $\chi^{\,\xi}$ and using the
definition of $D^L/Dt$ in equation (\ref{u-Xi-rel}) yields the {\bf advective
derivative relation},
\begin{equation}\label{advect-der-eqn}
\Big(\frac{\partial\chi}{\partial t}\Big)^\xi
+
T\chi\cdot\frac{D^L}{Dt}\Xi\,(\mathbf{x},t)
=
\Big(\frac{\partial\chi}{\partial t}
+
T\chi\cdot\mathbf{u}\Big)^\xi
\,,
\end{equation}
so ${D^L\chi^\xi}/{Dt}=({D\chi}/{Dt})^\xi$.
As in equation (\ref{LA-vee}) for the velocity, the {\bf Lagrangian
mean} $\bar\chi^L$ of a fluid quantity $\chi$ is defined as
\begin{equation}\label{Lag-mean-def}
\bar\chi^L(\mathbf{x},t)
\equiv
\overline{\chi^\xi(\mathbf{x},t)}
=
\overline{\chi(\mathbf{x}^\xi,t)}
=
\overline{\chi\big(\,g(t)\cdot\mathbf{x}_0,t\big)}
\,.
\end{equation}
Taking the Lagrangian mean of equation (\ref{advect-der-eqn}) and 
again  using its projection property yields
$ 
\dot{\bar\chi}^L
=
{D^L\bar\chi^L}/{Dt}
=
\overline{({D\chi}/{Dt})}^L
=
\bar{\dot{\chi}}^L
.$ 
Thus, the Lagrangian mean defined in (\ref{Lag-mean-def}) commutes with
the material derivative. 

\noindent
{\bf Transformation factors of advected quantities.} 
Advective transport by $g(t)$ and $\tilde{g}(t)$ is defined by
\begin{eqnarray*}
a(\mathbf{x}^\xi,t)
=
a_0\cdot g^{-1}(t)
\quad\hbox{and}\quad
\tilde{a}(\mathbf{x},t)
=
a_0\cdot \tilde{g}^{-1}(t)
\,,
\end{eqnarray*}
where $a_0=a(\mathbf{x}_0,0)=\tilde{a}(\mathbf{x}_0,0)$,
with $a,\tilde{a}\in V^*$ and the factorization $g(t)=\Xi\,(t)\cdot
\tilde{g}(t)$ implies 
$
\tilde{a}(\mathbf{x},t)=a\cdot\,\Xi\,(\mathbf{x},t).
$
Since $a$ and $\tilde{a}$ refer to the same initial conditions, $a_0$, 
\begin{equation}\label{eff-def}
a_0\!\cdot \tilde{g}^{-1}(t)
=
\tilde{a}(\mathbf{x},t)
=
a\cdot\,\Xi\,(\mathbf{x},t)
\equiv
{\cal F}(\mathbf{x},t)\!\cdot\! a^\xi(\mathbf{x},t)
\,.
\end{equation}
Note that the right side of this equation is potentially rapidly
varying, but the left side is a mean advected quantity. 
Here ${\cal F}(\mathbf{x},t)$ is the {\bf tensor transformation factor}
of the advected quantity $a$ under the change of variables
$\Xi\,:\mathbf{x}\to\mathbf{x}^\xi$. For example, the density, $D$,
transforms as
\begin{eqnarray}\label{D-Xform}
D^\xi\det(T\Xi)(\mathbf{x},t) 
=
\tilde{D}(\mathbf{x},t)
\,,\quad
&&
{\cal F}(\mathbf{x},t) 
= \det(T\Xi)
\,,
\\
\hbox{and $\tilde{D}$ advects as }
\hspace{22mm}&&
{\partial_t}\tilde{D}
=-\,
{\rm div}(\tilde{D}\tilde{\mathbf{u}})
\,.
\label{D-advect}
\end{eqnarray}
The transformation factors are 1, $\det(T\Xi)$ and
$\mathsf{K}\equiv\det(T\Xi\,)\,T\Xi^{\,-1}$, for an
advected scalar, density and vector field, respectively.
In each case, the corresponding transformation factor ${\cal F}$ appears
in a    {\bf variational relation} for an advected quantity, expressed
via equation (\ref{eff-def}) as
\begin{equation}\label{var-a-rel}
\delta{a}^\xi
=
\delta\,({\cal F}^{-1}\cdot\tilde{a})
=
{\cal F}^{-1}\cdot\delta\tilde{a}
+
(\delta{\cal F}^{-1})\cdot\tilde{a}
\,.
\end{equation}
This formula will be instrumental in establishing the main result.

\noindent
{\bf Lagrangian averaged Euler-Poincar\'e Theorem.} 
Let the following list of assumptions hold
\cite{HMR[1998a],HMR[1998b]}.

$\bullet\quad$ There is a {\it right\/} representation of Lie
group $G$ on the vector space $V$ and $G$ acts in the natural way on
the {\it right\/} on $TG \times V^\ast$: $(v_g, a)h = (v_gh, ah)$.

$\bullet\quad$ The function $ L : T G \times V ^\ast
\rightarrow \mathbb{R}$ is right $G$--invariant.

$\bullet\quad$ In particular, if $a_0 \in V^\ast$, define the
Lagrangian $L_{a_0} : TG \rightarrow \mathbb{R}$ by
$L_{a_0}(v_g) = L(v_g, a_0)$. Then $L_{a_0}$ is right
invariant under the lift to $TG$ of the right action of
$G_{a_0}$ on $G$, where $G_{a_0}$ is the isotropy group of $a_0$.

$\bullet\quad$  Right $G$--invariance of $L$ permits one to define
$\ell: {\mathfrak{g}} \times V^\ast \rightarrow \mathbb{R}$ by
$ 
\ell(v_gg^{-1}, a_0g^{-1}) = L(v_g, a_0).
$ 
Conversely,  this relation defines for any
$\ell: {\mathfrak{g}} \times V^\ast \rightarrow
\mathbb{R} $ a right $G$--invariant function
$ L : T G \times V ^\ast
\rightarrow \mathbb{R} $.

$\bullet\quad$ For a curve $g(t) \in G, $ let
$ 
u(t) \equiv \dot{g}(t) g(t)^{-1}\in TG/G\cong \mathfrak{g}
$ 
and define the curve $a(t)$ as the unique solution of the linear
differential equation with time dependent coefficients 
$ 
\dot a(t) = -a(t)u(t)
$ 
where the action of $u\in\mathfrak{g}$ on the initial condition $a(0)
= a_0\in V^*$ is denoted by concatenation from the right. This
solution can be written as the {\bf advective
transport relation},
$ 
a(t)
= a_0g(t)^{-1}.
$ 

$\bullet\quad$ The GLM factorization holds, 
$g(t)=\Xi\,(t)\cdot\tilde{g}(t)$ with
$\bar{g}(t) = \overline{\Xi\,(t)\cdot \tilde{g}(t)}
=\tilde{g}(t)$ and $\bar{\tilde{g}}(t)=\tilde{g}(t)$. 
\smallskip 

\noindent
{\bf LAEP Theorem.}\par\noindent
{\it 
\noindent 
The following are equivalent:

\noindent
$\overline{\bf i}$ The averaged Hamilton's principle holds
\begin{equation} \label{Avg-Ham-Princ}
\delta \int _{t_1} ^{t_2} 
\overline{
L_{a_0}(g(t), \dot{g} (t))
}
\, dt 
=
0
\end{equation}
for variations $\delta g(t)$
of $ g (t) $ vanishing at the endpoints.

\noindent
$\overline{\bf ii}$   The averaged Euler--Lagrange
equations for $\bar{L}_{a_0}$ are satisfied on $T^*\tilde{G}$,
\begin{equation} \label{Avg-EL-eqns}
\overline{
\frac{\delta L_{a_0}}{\delta {g}}\cdot T\Xi\,
}
-
\overline{
\frac{d}{dt}
\frac{\delta L_{a_0}}{\delta \dot{g}}
\cdot T\Xi
}
=
0
\end{equation}
%

\noindent
$\overline{\bf iii}$   The averaged constrained variational
principle
\begin{equation} \label{Avg-var-princ}
\delta \int _{t_1} ^{t_2}  
\overline{
\ell\,\big(u(t), a(t)\big)
} dt = 0
\end{equation}
holds, using variational relations of the form
\begin{eqnarray} \label{delta-you}
\delta u
&=&
\big(\partial_t + {\rm ad}_u\big)\eta^{\,\prime}
+
\big(
T\Xi\,\cdot
\big(\partial_t + {\rm ad}_{\tilde{u}}\big)\tilde{\eta}
\,\big)\,\Xi^{\,-1}
\nonumber\\
\delta a
&=&
-\,
a\,\eta
=
\delta\big({\cal F}^{\,-1}\cdot\tilde{a}\big)\,\Xi^{\,-1}
\nonumber\\
&=&
-\,
a\,\eta^{\,\prime}
-\,\big(
{\cal F}^{\,-1}\!\cdot(\tilde{a}\,\tilde{\eta}\,)\big)\,\Xi^{\,-1}
\,,
\label{delta-ay}
\end{eqnarray}
where Lie derivatives of advected quantities by the vector fields
$\eta^{\,\prime}(t)\equiv
\delta\,\Xi\,\, \Xi^{\,-1}$,
$\tilde{\eta}(t) \equiv \delta \tilde{g}\,\tilde{g}^{-1}$ 
and 
\begin{equation} \label{eta-full}
\eta
\equiv
\delta{g}\, g^{-1}
=
\eta^{\,\prime}
+
(T\Xi\,\cdot\,\tilde{\eta}\,)\,\Xi^{\,-1}
\,,
\end{equation}
are indicated by concatenation on the right and these three vector
fields all vanish at the endpoints.

\noindent
$\overline{\bf iv}$
The Euler--Poincar\'e (EP) equation holds on
${\mathfrak{g}} \times {V}^\ast$
\begin{equation} \label{EP-eqn-right}
\Big(
\frac{ \partial}{\partial t}
+
 {\rm ad}_{u}^{\ast} 
\Big)
\frac{\delta \ell}{\delta u} 
= 
\frac{\delta \ell}{\delta a} \diamond a
\,,\end{equation}
and the Lagrangian averaged Euler--Poincar\'e (LAEP)
equation holds on $\tilde{\mathfrak{g}} \times \tilde{V}^\ast$
\begin{equation} \label{LAEP-eqn-right}
\Big(\frac{ \partial}{\partial t} 
+
 {\rm ad}_{\tilde{u}}^{\ast} \Big)
\overline{\Big(
\frac{\delta \ell}{\delta u^{\,\xi}}
\cdot T\Xi\Big)}
=
\overline{\Big(
\frac{\delta \ell}{\delta a^{\,\xi}} 
\cdot {\cal F}^{-1}\Big)}
\diamond \tilde{a}
\,.
\end{equation}
}

\noindent
{\bf Notation.}
In equations (\ref{EP-eqn-right}) and (\ref{LAEP-eqn-right}), the
operations  ad${^*}$ and $\diamond$ are defined by using the
$L_2$ pairing  $\langle{f,g}\rangle=\int fg\, d^3x$. The ad$^*$
operation is defined as (minus) the $L_2$ dual of the Lie algebra
operation, ad, or commutator, ad$_u\,\eta=-\,[u,\eta]$, for vector
fields,
$ 
-\,
\big\langle
{\rm ad}^*_u\, \mu, \eta
\big\rangle
\equiv
\big\langle
\mu, {\rm ad}_u\,\eta
\big\rangle
$. 
The diamond operation $\diamond$ is defined as (minus) the $L_2$ dual
of the Lie derivative, namely, 
$ 
\big\langle b\diamond a\,,\, \eta \big\rangle
\equiv
-
\big\langle b\,,\, {\pounds_\eta} a\big\rangle
=
-
\big\langle b\,,\, a\,\eta\big\rangle
,$ 
where ${\pounds_\eta}  a$ denotes the Lie derivative with respect to
vector field $\eta$ of the tensor $a$, and $a$ and $b$ are dual
tensors.\smallskip

\noindent
{\bf Proof of the LAEP Theorem.}
The equivalence of $\overline{\bf i}$ and $\overline{\bf ii}$ holds
for any configuration manifold, so it holds again here. To compute
the averaged Euler-Lagrange equation (\ref{Avg-EL-eqns}), we use the
following {\bf variational relation} obtained from the composition of
maps $g(t)=\Xi\,(t)\cdot
\tilde{g}(t)$, cf. the velocity relation (\ref{vee-rel}),
\begin{equation}\label{var-gee-rel}
\delta{g}(t)
=
\delta{\Xi}\,(t)\cdot\tilde{g}(t)
+
T\Xi\,(t)\cdot \delta\tilde{g}(t)
\,.
\end{equation}
Hence, we find
\begin{eqnarray} \label{Avg-Ham-Princ-proof}
0 
&=& 
\delta \int _{t_1} ^{t_2} 
\overline{
L_{a_0}(g(t), \dot{g} (t))
}
\, dt 
\nonumber\\
&=&
\int _{t_1} ^{t_2} 
\bigg(\
\overline{
\frac{\delta L_{a_0}}{\delta {g}}\cdot \delta {g} 
}
+
\overline{
\frac{\delta L_{a_0}}{\delta \dot{g}}\cdot \delta \dot{g} 
}\
\bigg)
\,dt
\nonumber\\
&=&
\int _{t_1} ^{t_2} 
\overline{
\bigg(\!
\Big(\,
\frac{\delta L_{a_0}}{\delta {g}}
-
\frac{d}{dt}
\frac{\delta L_{a_0}}{\delta \dot{g}}
\Big)\cdot\delta\,\Xi\,(t)
\!\bigg)
}
\cdot\tilde{g}\,dt
\\
&&
+
\int _{t_1} ^{t_2} 
\bigg(\
\overline{
\frac{\delta L_{a_0}}{\delta {g}}\cdot T\Xi\,
}
-
\overline{
\frac{d}{dt}
\frac{\delta L_{a_0}}{\delta \dot{g}}
\cdot T\Xi}\
\bigg)\cdot\delta\tilde{g}\,dt
\,.
\nonumber
\end{eqnarray}
In the last equality, the first integral vanishes for any $\delta\,\Xi$,
thus ensuring that the Euler-Lagrange equations are satisfied
{\it before} averaging is applied. Vanishing of the second integral for
arbitrary $\delta\tilde{g}$ then yields the averaged Euler-Lagrange
equations (\ref{Avg-EL-eqns}).

The equivalence of $\overline{\bf iii}$ and $\overline{\bf iv}$
in the LAEP theorem now follows by substituting the variations
(\ref{delta-ay}) into (\ref{Avg-var-princ}), and integrating by parts,
to find 
\begin{eqnarray}\label{EP-derivation-proof}
0 
&=&
\delta \int_{t_1}^{t_2}
   \overline{\ell(u, a)}\, dt 
=
\int_{t_1}^{t_2}
\Big\langle
\overline{
\frac{\delta \ell}{\delta u}\,,\,\delta u 
}
\Big\rangle
+
\Big\langle
\overline{
\frac{\delta \ell}{\delta a}\,,\,\delta a 
}
\Big\rangle
\,dt 
\nonumber \\
  &=& -\
\int_{t_1}^{t_2}
\Big\langle
\overline{
\big(\partial_t + {\rm ad}^*_u\big)
\frac{\delta \ell}{\delta u}
-
\frac{\delta \ell}{\delta a}
\diamond a
\,,\,
\eta^{\,\prime}
}
\Big\rangle
\,dt
\nonumber\\
  &&
 -\int_{t_1}^{t_2}
 \bigg\langle
\Big(\frac{ \partial}{\partial t} 
+
 {\rm ad}_{\tilde{u}}^{\ast} \Big)
\overline{\Big(
\frac{\delta \ell}{\delta u^\xi}
\cdot T\Xi\Big)}
\\
  &&\hspace{25mm}-\
\overline{\Big(
\frac{\delta \ell}{\delta a^\xi} 
\cdot {\cal F}^{-1}\Big)}
\diamond \tilde{a}
\,,\,
\tilde{\eta}
\bigg\rangle
\,dt
\,.
\label{LAEP-derivation-proof}
\end{eqnarray}
Thus, the independent variations
$\eta^{\,\prime}$ in (\ref{EP-derivation-proof}) and $\tilde{\eta}$ in
(\ref{LAEP-derivation-proof}) result in the EP motion equation
(\ref{EP-eqn-right}) and the LAEP motion equation
(\ref{LAEP-eqn-right}), respectively.

Finally we show that $\overline{\bf i}$ and $\overline{\bf iii}$ are
equivalent. First note that the $G$--invariance of $L:TG
\times V^\ast \rightarrow \mathbb{R}$ and the definition of 
$a(t) = a_0g(t)^{-1}$ imply that the
integrands in (\ref{Avg-Ham-Princ}) and
(\ref{Avg-var-princ}) are equal. In fact, this holds both before and
after averaging. Moreover, all variations $\delta g(t) \in TG$ of
$g(t)$ with fixed endpoints induce and are induced by variations
$\delta u(t) \in \mathfrak{g}$ of $u(t)$ of the form $\delta u = \partial\eta
/\partial t + {\rm ad}_u\,\eta $ with $\eta(t) \in \mathfrak{g}$
vanishing at the endpoints. The relation between $\delta g(t)$ and
$\eta(t)$ is given by $\eta(t) = \delta g(t)g(t)^{-1}$. The
corresponding statements also hold for the prime- and tilde-variables in
the variational relations (\ref{delta-ay}) that
are used in the calculation of the other equivalences.
This finishes the proof of the LAEP Theorem.

\noindent
{\bf Lie derivative vs ad$^*$.}
The equality ad$^*_u\mu={\pounds_u}\mu$ holds for any one-form
density $\mu$ (such as $\mu=\delta\ell/\delta u$, the variational
derivative). Thus, the LAEP motion equation (\ref{LAEP-eqn-right}) may be
written equivalently using Lie derivatives as\vspace{-3mm}
\begin{eqnarray*}
\hspace{-10mm}
\Big(
\frac{ \partial}{\partial t}
+
{\pounds_{\tilde{u}}}
\Big)
\overline{\Big(
\frac{\delta \ell}{\delta u^\xi}
\cdot T\Xi\Big)}
=
\overline{\Big(
\frac{\delta \ell}{\delta a^\xi} 
\cdot {\cal F}^{-1}\Big)}
\diamond \tilde{a}
\,.
\end{eqnarray*}
%
In this notation, the advection
of mass takes the form 
$({\partial_t} + {\pounds_{\tilde{u}}}) \tilde{D} = 0 $
and one immediately obtains the 

\noindent
{\bf LA Kelvin-Noether Circulation Theorem.} \par\noindent
{\it  On defining
$\tilde{v}\equiv 
\overline{\Big(
\frac{\delta \ell}{\delta u^\xi}
\cdot T\Xi\Big)}/{\tilde{D}}
$, 
\begin{eqnarray*}
\frac{d}{dt}
\oint_{c(\tilde{u})}
\tilde{v}
=
\oint_{c(\tilde{u})}
\frac{1}{\tilde{D}}\
\overline{\Big(
\frac{\delta \ell}{\delta a^\xi} 
\cdot {\cal F}^{-1}\Big)}
\diamond \tilde{a}
\,,
\nonumber
\end{eqnarray*}
for any closed curve $c(\tilde{u})$ following the LA fluid motion.
}

\noindent
The same result is obtained by applying LA directly to Kelvin's
Theorem for the EP motion equation (\ref{EP-eqn-right}).


\noindent
{\bf Applying the LAEP theorem to incompressible fluids.}
The Lagrangian averaged Euler (LAE) equations for an incompressible fluid are
derived from the LAEP theorem using the reduced averaged
Lagrangian,
\begin{equation} \label{reduced-avg-spatial-lag}
\bar\ell
= 
\int d^3x\
\Big[\
\frac{1}{2}\,\tilde{D}\,
\overline{ \,|\mathbf{u}^\xi|^2 }
+
\overline{p^\xi\,
\Big(\det T\Xi\,-\tilde{D}\Big)
}\
\Big]
\,.
\end{equation}
The pressure constraint implies that the mean advected density is related to
the mean fluid trajectory by 
$ 
\tilde{D} = \overline{ \,\det T\Xi\,}
\,.
$ 
Thus, in general, the LAE fluid velocity has a
nonzero divergence \cite{AM[1978]}, since (\ref{D-advect}) for $\tilde{D}$
implies
\begin{equation} 
{\rm div}\,\tilde{\mathbf{u}}
=
-\,
\frac{1}{\tilde{D}}
\Big(\frac{\partial}{\partial t}
+
\bar{\mathbf{u}}^L\cdot\nabla\Big)\tilde{D}
\ne0
\,.
\nonumber
\end{equation}
In principle, one may restrict $g(t)$ and both its
factors $\Xi(t)$ and $\tilde{g}(t)$ to the space of volume-preserving
diffeomorphisms, for which $\det T\Xi\equiv1$. However, 
some LA processes may not respect this restriction and, in general,
${\rm div}\,\tilde{\mathbf{u}}\ne0$. In \cite{MRS[2001],MS[2001]}, though,
${\rm div}\,\tilde{\mathbf{u}}=0$ is accomplished.

For $\bar\ell$ in (\ref{reduced-avg-spatial-lag}) the LAEP equation
(\ref{LAEP-eqn-right}) gives 
\begin{equation} \label{LAEP-eqns}
\frac{ \partial}{\partial t} \tilde{v}_i
+
\tilde{u}^j\frac{ \partial}{\partial x^j}\tilde{v}_i
+
\tilde{v}_j\frac{ \partial}{\partial x^i}\tilde{u}^j
+
\frac{ \partial}{\partial x^i}\tilde{\pi}
= 
0
\,,\end{equation}
with mean fluid quantities $\tilde{v}_i$ and $\tilde{\pi}$ defined as
\begin{eqnarray*} \label{LAEP-defs}
\tilde{v}_i
=
\frac{1}{\tilde{D}}
\frac{\delta \ell}{\delta \tilde{u}^{\,i}}
= 
\overline{u^{\,\xi}_j(T\Xi)^j_i}
\,,\quad
\tilde{\pi}=-\,\frac{\delta \ell}{\delta \tilde{D}}
=
-\,\frac{1}{2}\,\overline{\, |\mathbf{u}^{\,\xi}|^2} + \bar{p}^L
\,.
\end{eqnarray*}
When $T\Xi = Id + \nabla \xi$, for a vector field
$\xi=\mathbf{x}^\xi-\mathbf{x}$, one finds 
$ 
\tilde{\mathbf{v}}
= 
\overline{\mathbf{u}^\xi}
+
\overline{\frac{D^L}{Dt}\xi_j\nabla\xi^j}
\equiv
\bar{\mathbf{u}}^L - \bar{\mathbf{p}}
\,.
$ 
The term $\bar{\mathbf{p}}$ is called the pseudomomentum
\cite{AM[1978]}. See, e.g., \cite{GH[1996],Holm[1999],Holm[2001]} for
discussions of the role of pseudomomentum in GLM theory.

\noindent
{\bf Momentum balance.}\par\noindent
{\it 
Following the EP theory of Holm, Marsden {and} Ratiu
\cite{HMR[1998a],HMR[1998b]} leads to momentum balance for the LAE
equations,
\begin{equation} \label{LAEP-mom-conserv}
\frac{ \partial}{\partial t} (\tilde{D}\tilde{v}_i)
+
\frac{ \partial}{\partial x^j}\Big(
\tilde{D}\tilde{v}_i\tilde{u}^j
+
\bar{p}^L\delta^j_i\Big)
= 
\frac{\tilde{D}}{2}
\frac{ \partial\overline{\,|\mathbf{u}^\xi|^2}}{\partial
x^i}\,\bigg|_{exp}\,,
\end{equation}
where subscript $exp$ refers to the {\bf explicit} spatial
dependence that yields a mean force arising from the $\Xi-$terms in
$\overline{\,|\mathbf{u}^\xi|^2}
=
\overline{\,|D^L\,\Xi\,/Dt\,|^2}$ that appear in equation
(\ref{u-xi-def}).
}

\noindent
{\bf Recent progress toward closure.}
Of course, the LAE equations (\ref{LAEP-eqns}) are not yet closed. As
indicated in their momentum balance relation (\ref{LAEP-mom-conserv}),
they depend on the unspecified Lagrangian statistical properties
appearing as the $\Xi-$terms in the definitions of $\tilde{\mathbf{v}}$
and $\tilde{\pi}$. Until these properties are modeled or prescribed,
the LAE equations are incomplete.

Progress in formulating and analyzing a closed system of fluid equations
related to the LAE equations has recently been made in the EP context.
Such closed model LAE equations were first obtained in Holm, Marsden
{and} Ratiu \cite{HMR[1998a],HMR[1998b]}.  For more discussion of this
type of equation and its recent developments as a turbulence model, see
papers by Chen {\it et al.}
\cite{Chen-etal[1998],Chen-etal[1999a],Chen-etal[1999b],Chen-etal[1999c]},
Shkoller \cite{Shkoller[1998]},  Foias {\it et al.} \cite{FHT[1999],FHT[2001]},
Marsden, Ratiu {and} Shkoller \cite{MRS[2001]} and Marsden {and} Shkoller
\cite{MS[2001]}.  A self-consistent variant of the LAE closure was
also introduced in Gjaja {and} Holm \cite{GH[1996]} in the Lagrangian fluid
specification, see Holm \cite{Holm[1999],Holm[2001]} for further discussion of
that approach. 

The LAEP approach is versatile enough to derive LA equations
for {\it compressible} fluid motion, as well. This was already shown in the
original GLM theory \cite{AM[1978]}. For brevity, we
only remark that the LAEP approach also preserves helicity conservation for
compressible flows and it preserves magnetic helicity and cross-helicity
conservation when applied to magnetohydrodynamics (MHD). For more details in
this regard, see \cite{Holm[2001]}.

{\bf Acknowledgments.} I am grateful for stimulating discussions of this topic
with P. Constantin, G. Eyink, U. Frisch, J. E. Marsden, M. E. McIntyre, I.
Mezic, S. Shkoller and A. Weinstein. Some of these discussions took
place at Cambridge University while the author was a visiting professor
at the Isaac Newton Institute for Mathematical Science. This work was
supported by the US DOE under contracts
W-7405-ENG-36 and the Applied Mathematical Sciences Program KC-07-01-01.


\end{multicols}

\end{document}